\documentclass[floatfix,twocolumn,aps,prl,showpacs,amsmath,amssymb]{revtex4}
\usepackage{epsfig}
\usepackage[latin1]{inputenc}

\usepackage[dvips]{color}

\begin{document}

\title{Bosonization of Weyl Fermions and Free Electrons}

\author{E. C. Marino}
\affiliation{Instituto de F\'\i sica, Universidade Federal do Rio de Janeiro, C.P.68528, Rio de Janeiro RJ, 21941-972, Brazil }

\date{\today}

\begin{abstract}
The electron, discovered by Thomson by the end of the nineteenth century, was the first experimentally observed particle.
The Weyl fermion, though theoretically predicted since a long time, was observed in a condensed matter
environment in an experiment reported only a few weeks ago. Is there any linking thread connecting the first and the last observed fermion (quasi)particles?
The answer is positive. By generalizing the method known as bosonization, the first time in its 
full complete form, for a spacetime with 3+1 dimensions,
we are able to show that both electrons and Weyl fermions can be expressed in terms  of the same boson field, namely
the Kalb-Ramond anti-symmetric tensor gauge field. The bosonized form of the Weyl chiral currents lead to the angle-dependent magneto-conductance behavior  observed
in these systems.
\end{abstract}

\pacs{}

\maketitle

{\textbf{I.\,Introduction}}

Nature exhibits two completely different classes of particles: fermions and bosons. Apparently all matter 
is ultimately composed out
of fermion particles, namely, electrons, quarks and then, protons, neutrons and so on. The interactions
among these matter particles, conversely, seem to be mediated only by bosonic particles, such as photons, gluons
and massive vector mesons, which are the quantum-mechanical manifestation of the gauge fields that 
intermediate such interactions.
Is it possible to bridge the gap separating the fermionic matter constituents from the bosonic conveyors of their
interactions?
Is it conceivable that a super-unification of matter with its interactions could be achieved?

In this work, we generalize the method of complete bosonization (bosonization of the fermion field 
itself) to four-dimensional spacetime, thereby demonstrating that both Dirac (electrons) and Weyl fermions can
be expressed in terms of a 2-tensor, Kalb-Ramond bosonic gauge field. The fermion particles then appear as
topological excitations of the gauge field. 

The result has far reaching consequences. In the condensed matter framework, our results apply to the
Weyl fermion quasiparticles, experimentally observed very recently by photoemission spectroscopy methods in
the $TaAs$ and $NbAs$ semimetals \cite{weyl}. In particle physics, conversely, 
it bears on the very nature of elementary point particles such as the electron. Take, for instance QED, 
which results from the interplay of the bosonic gauge vector 
electromagnetic field with its charge bearing fermion sources such as electrons, for instance. With the full electron bosonization, one 
would be able to have QED formulated as a theory involving just gauge fields describing both electrons and photons.

Bosonization is an extremely powerful method by which fermionic fields are mapped into bosonic ones.
It was firstly developped in one spatial dimension \cite{bos1d,bos1d1,sm}, where it found important applications ranging from
condensed matter to particle physics.
The Tomonaga-Luttinger \cite{tl} system 
is an example of the former, while QED$_2$ of massless
fermions \cite{ls} is an example of the latter. The method of bosonization has led, in the first case, to the obtainment of the
exact energy spectrum, correlations functions, response functions, etc. of such strongly interacting nonlinear systems.
In the second case, conversely, it produced and exact operator solution of QED$_2$, which provided a deep understanding of general features of QCD$_4$,
such as color confinement/screening, chiral symmetry breaking, topological vacua, instantons and the hadronic spectrum
within an exactly soluble, completely controlable system.

The excellence of the bosonization method has made desirable its generalization to higher dimensions. Early attempts in this direction were made in \cite{3,4,5}. Later developments can be found in \cite{bos4d1,bos4d,bos4d2}.
Nevertheless, a complete bosonization of fermions, by which we mean the full expression, not only of the lagrangean and currents, but of the fermion field itself
in terms of a bosonic one, was achieved for the first time for $D>2$ in \cite{em1}. There a
complete bosonization of the masless free Dirac field in D=2+1 was firstly obtained, in terms of a bosonic vector gauge field.

It became clear, then, that the full bosonization could be extended to higher dimensions, nevertheless, it would be 
an exact mapping only for non-interacting theories. This is the main difference between bosonization in higher D,
when compared to bosonization in D=1+1. There exists, however, a common feature among the bosonization procedures in any 
spacetime dimension. This is an underlying order-disorder duality structure, which has its roots in
statistical mechanics \cite{odd}, out of which emerge certain  dual, order and disorder operators \cite{odd,odd1}.

Bosonization relies on the fact that a fermion field may be expressed
as a product of such order and disorder operators, \cite{em2,fm,em1} defined in terms of a bosonic field.
In this framework, one of the dual operators carries the topological charge of the bosonic field, which
is identically conserved, whereas the other carries a charge, which is conserved by means of the the bosonic field equation.
Through bosonization, then, the topological charge of the bosonic theory, carried by one of the dual operators, 
is identified with the fermion charge, while the charge carried by the dual companion becomes the fermion chirality
\cite{em1,em2}.

{\textbf{II.\,Weyl and Dirac Fermions }}

A Dirac Fermion decomposes in two-components Weyl fermions, such that
\begin{eqnarray}
\psi &=& \left ( \begin{array}{l}           
                                    \psi_L \\
                                    \psi_R
                                        \end{array} 
                              \right )   \ \ \
\label{677a}
\end{eqnarray}
when  we use the Weyl representation of the Dirac matrices.
The lagrangean of a massless Dirac field, in this case, reads
\begin{eqnarray}
\mathcal{L}= i \bar{\psi} \gamma^\mu \partial_\mu \psi = \psi_L^\dagger \sigma^\mu \partial_\mu \psi_L
+ \psi_R^\dagger \bar\sigma^\mu \partial_\mu \psi_R
\label{675a}
\end{eqnarray}
where
$\sigma^\mu = (\mathbb{I}, \sigma^i) $ and $  \bar \sigma^\mu = (\mathbb{I}, -\sigma^i)$ and $\mathbb{I}$ is the rank-2 identity matrix.

A mass term for the Dirac field is given by
\begin{eqnarray}
\mathcal{L}_M=M \bar{\psi} \psi =M\left[ \psi_L^\dagger \psi_R + \psi_R^\dagger \psi_L\right]
\label{675ab}
\end{eqnarray}
The current and chiral current, accordingly,  will be given respectively by
\begin{eqnarray}
\bar{\psi} \gamma^\mu \psi = \psi_L^\dagger \sigma^\mu \psi_L + \psi_R^\dagger \bar\sigma^\mu \psi_R
\nonumber \\
\bar{\psi}\gamma^5 \gamma^\mu  \psi = \psi_L^\dagger \sigma^\mu \psi_L - \psi_R^\dagger \bar\sigma^\mu \psi_R
\label{67ab}
\end{eqnarray}

{\textbf{III.\,Pre-Bosonization}}

It is convenient to express the energy and momentum in terms of the rapidity variable $\chi\in[0,\infty)$, such that
 in the positive energy, time-like region of Minkowski space, we have (in other regions there will be corresponding
expressions)
\begin{eqnarray}
k^0=k \cosh\chi \ \ \ ;\ \ \  |\textbf{k}|= k \sinh \chi
\label{678a}
\end{eqnarray}
where $k=\sqrt{k_\mu k^\mu}$.

We then have,
\begin{eqnarray}
\sigma^\mu k_\mu = k\left [ \mathbb{I} \cosh\chi + \hat{r}\cdot\mathbf{\sigma} \sinh\chi \right ]                    
\label{679a}
\end{eqnarray}
where $\hat r$ is the radial unit vector of the spherical coordinate system.

Before bosonizing the Weyl fermion fields $\psi_L$ and $\psi_R$, we introduce the new spinor fields
\begin{eqnarray}
\Psi_L &=& T_L \psi_L\ \ \ \ \ \ \ \Psi_R= T_R \psi_R
\nonumber \\
T_{L,R} &=& 
\left [ \mathbb{I} \cosh\frac{\chi}{2} \pm \hat{\varphi}\cdot\mathbf{\sigma} \sinh\frac{\chi}{2} \right ]
\left [ \mathbb{I} -i \hat{\theta}\cdot\mathbf{\sigma} \right ]                    
\label{680a}
\end{eqnarray}
such that
\begin{eqnarray}
\psi^\dagger_{A} \sigma^\mu k_\mu \psi_{A}&=&\Psi^\dagger_{A} k \mathbb{I}\Psi_{A} \ \ \ ; \ \ \ A=L,R
\label{681a}
\end{eqnarray}
The canonical transformations $T_{L,R}$, render the massless lagrangean diagonal. It is similar to the Foldy-Wouthuysen transformation but is not unitary. 

From the previous equation, we conclude
the $\Psi$-field euclidean correlation functions are
\begin{eqnarray}
\langle\Psi_L(x)\Psi^\dagger_L(y)\rangle &=& \langle\Psi_R(x)\Psi^\dagger_R(y)\rangle=\frac{\mathbb{I}}{2\pi^3|x-y|^3}
\nonumber \\
\langle\Psi_L(x)\Psi^\dagger_R(y)\rangle &=& \langle\Psi_R(x)\Psi^\dagger_L(y)\rangle=0
\label{684a}
\end{eqnarray}

Considering that $T^\dagger_{L}T_{R} = T^\dagger_{R}T_{L}=\mathbb{I}$, it follows that the mass term
becomes
\begin{eqnarray}
\mathcal{L}_M=M \bar{\psi} \psi =M\left[ \Psi_L^\dagger \Psi_R + \Psi_R^\dagger \Psi_L\right]
\label{675abc}
\end{eqnarray}

{\textbf{IV.\, Lagrangean and Current Bosonization}}

Consider the generating functional of current correlation functions 
\begin{eqnarray}
Z[J]&=&\int D\psi D\bar \psi \exp \left \{- \int d^4z [i \bar\psi \not\!\partial \psi
-J_\mu j^\mu]\right \} =
\nonumber \\
 &=&Z_2[J]Z_{N>2}[J]
\nonumber \\
Z_2[J]&=&  \exp \left \{\frac{1}{2} \int d^4z J_\mu \Pi_1^{\mu\nu} J_\nu\right \} 
\label{634aa}
\end{eqnarray}
where $\Pi_1^{\mu\nu}$ is the one-loop, massless vacuum polarization tensor and $Z_{N>2}[J]$ contains only higher-than-quadratic 
terms.
It follows that
the fermion current $j^\mu = \bar\psi \gamma^\mu \psi $ two-point correlation function in momentum space is
\begin{eqnarray}
\langle j^\mu j^\nu\rangle(k) = \Pi_1^{\mu\nu}(k) = \frac{1}{24\pi^2}\left[k^2\delta^{\mu\nu}-k^\mu k^\nu\right]
\label{635aa}
\end{eqnarray}
Then
assuming
the fermionic current $j^\mu = \bar\psi \gamma^\mu \psi $ is expressed as 
$$
j^\mu = \frac{1}{2}K^{\mu\alpha\beta} \left[B^L_{\alpha\beta}+B^R_{\alpha\beta} \right]
$$
in terms of the chiral bosonic 2-tensor fields $B^{L,R}_{\mu\nu}$, where the tensor $K^{\mu\alpha\beta}$
is to be determined, 
we may write the relevant generating functional as
\begin{eqnarray}
&\ &Z_2[J]=\int DB^L_{\mu\nu}DB^R_{\mu\nu} \times
\nonumber \\
&\ &\exp \left \{- \int d^4z \sum_{A=L,R}\left[\frac{1}{24} H^A_{\mu\nu\alpha} H_A^{\mu\nu\alpha} 
-\frac{1}{2} J_\mu K^{\mu\alpha\beta} B^A_{\alpha\beta}\right]\right \}
\nonumber \\
\label{636aa}
\end{eqnarray}
where $H_{\mu\nu\alpha}=\partial_\mu B_{\nu\alpha}+\partial_\nu B_{\alpha\mu}+\partial_\alpha B_{\mu\nu}$.
Since it is a free theory, arbitrary $2n$-point correlation functions of the fields $\Psi_{L,R}$ will be products of (\ref{684a})
and, in order to reproduce them within the bosonic theory we do not have to go beyond $Z_2[J]$ in order to find
the bosonic lagrangean and current \cite{bm}.

From the expression above, we may infer the following bosonization formulas for the lagrangean and current
\begin{eqnarray}
i \bar\psi \not\!\partial \psi&=&\sum_{A=L,R}\frac{1}{24} H^A_{\mu\nu\alpha} H_A^{\mu\nu\alpha} 
\nonumber \\ 
\bar\psi \gamma^\mu \psi &=&\frac{1}{2} \sqrt{\frac{\Box}{24\pi^2}} 
\epsilon^{\mu\nu\alpha\beta}\partial_\nu \left[B^L_{\alpha\beta}+B^R_{\alpha\beta} \right]
\label{652aa}
\end{eqnarray}
For the axial current, we have, in the absence of an electromagnetic field
\begin{eqnarray}
\bar\psi \gamma^\mu \gamma^5\psi =\frac{1}{2} \sqrt{\frac{\Box}{24\pi^2}} 
\epsilon^{\mu\nu\alpha\beta}\partial_\nu \left[B^L_{\alpha\beta}-B^R_{\alpha\beta} \right] 
\label{653ac}
\end{eqnarray}

If there is an applied EM field $A_\mu$, however, the axial current will acquire a topological term \cite{gth}.
For the chiral, $L,R$ Weyl currents, according to (\ref{67ab}), we must have consequently
\begin{eqnarray}
j^\mu_L &=& \psi_L^\dagger \sigma^\mu \psi_L =
\frac{1}{2} \sqrt{\frac{\Box}{24\pi^2}} 
\epsilon^{\mu\nu\alpha\beta}\partial_\nu B^L_{\alpha\beta} + \frac{1}{2} I^\mu
\nonumber \\ 
j^\mu_R &=& \psi_R^\dagger \bar\sigma^\mu \psi_R =
\frac{1}{2} \sqrt{\frac{\Box}{24\pi^2}} 
\epsilon^{\mu\nu\alpha\beta}\partial_\nu B^R_{\alpha\beta} - \frac{1}{2}  I^\mu
\label{652ab}
\end{eqnarray}
and consequently, for the Dirac chiral current, $j^\mu_5$
\begin{eqnarray}
\bar\psi \gamma^\mu \gamma^5\psi =\frac{1}{2} \sqrt{\frac{\Box}{24\pi^2}} 
\epsilon^{\mu\nu\alpha\beta}\partial_\nu \left[B^L_{\alpha\beta}-B^R_{\alpha\beta} \right] + I^\mu
\label{652ac}
\end{eqnarray}
In the absence of an external EM field, the $I^\mu$ topological term just vanishes. When there is an
EM background field $A_\mu$, then \cite{gth}
\begin{eqnarray}
I^\mu=\frac{1}{4\pi^2}\epsilon^{\mu\nu\alpha\beta}A_\nu \partial_\alpha A_\beta
\label{652ad}
\end{eqnarray}
implying that
\begin{eqnarray}
\partial_\mu j^\mu_5 = \partial_\mu I^\mu = - \frac{1}{16\pi^2} F^{\mu\nu} \tilde  F^{\mu\nu}
\label{652ae}
\end{eqnarray}
where the last term is the Chern-Pontryagin topological charge density of the EM field,
namely, the chiral anomaly \cite{chi}.

{\textbf{V.\,Field Bosonization}}

We now consider the bosonization of the fields $\Psi_{L,R}$. For this purpose, the fundamental step is the
identification of the underlying duality structure and the relevant dual operators. The basic building blocks were introduced
in \cite{em3}:
\begin{eqnarray}
\mu(x) = \exp \left \{  ia \int_{-\infty}^x d \xi^\mu \Box^{-1/2}\epsilon^{\mu\nu\alpha\beta}
\partial_\nu B_{\alpha\beta}(\xi)\right \}.
\label{685a}
\end{eqnarray}
and
\begin{eqnarray}
\sigma(S(C_y)) &=& \exp \left \{ - i\frac{b}{2\pi \rho } \int_{S(C_y)} d^2 \xi^{\mu\nu}B_{\mu\nu}(\xi)\right \}.
\label{686a}
\end{eqnarray}
where
$a$ and $b$ are dimensionless real parameters and the Kalb-Ramond field $B_{\mu\nu}$ is either $L$ or $R$. $S(C_{\textbf{y}})$ is a surface having as boundary a 
circle of radius $\rho$, centered at the point $\textbf{y}$.

The $\mu$ operator creates quantum eigenstates of the topological charge of the Kalb-Ramond field, which
according to (\ref{652aa}), is proportional to the fermion charge.
The $S(C_{\textbf{y}})$ operator is nonlocal, in the sense it depends on a closed curve $C_{\textbf{y}}$, very much like the 
Wilson loop. It creates a quantum state bearing a flux of the (vector) source
of the Kalb-Ramond field, $Q^i=\partial_j H^{0ij}$, along that closed curve. 

Since we are bosonizing
a local field, however, we must take the local limit, where the curve $C_{\textbf{y}}$ shrinks to a point. Hence, in order to have
a nontrivial result, we  must consider the flux per unit length, by dividing the $b$ coefficient by the
string length. Thus the curve $C_{\textbf{y}}$ 
 is assumed to be a circle of infinitesimal radius $\rho$. Notice that $a$ and $b$ are dimensionless 
parameters to be determined.

Let us evaluate now the four-points mixed order-disorder correlation function in the framework of the bosonic 
Kalb-Ramond
theory given by (\ref{652aa}). This is quadractic and the order and disorder operators are exponentials of linear forms in the
2-tensor field. Hence, the functional integral leading to the correlation function can be straightforwardly 
performed, giving the result

\begin{eqnarray}
&\ &\langle \sigma(C_{x_1})\mu(x_2)\mu^\dagger(y_2)\sigma^\dagger(C_{y_1}) \rangle=
\nonumber \\ 
&\ &\exp \left \{-\frac{b^2}{2\pi^3}[ \ln \mu|x_1-y_1| - \ln \mu |\epsilon|]
\right .
\nonumber \\
&\ &\left .
-\frac{a^2}{2\pi^3}[ \ln \mu |x_2-y_2| - \ln \mu|\epsilon|]\right \}
\label{687aa}
\end{eqnarray}

In the limit $x_1\rightarrow x_2 =x$, $x_1\rightarrow x_2 =x$, we obtain
\begin{eqnarray}
&\ &\langle \sigma(C_{x_1})\mu(x_2)\mu^\dagger(y_2)\sigma^\dagger(C_{y_1}) \rangle\longrightarrow
\nonumber \\ 
&\ &\exp \left \{-\frac{a^2 + b^2}{2\pi^3}\left[ \ln \mu|x-y| - \ln \mu |\epsilon|\right] \right \}
\label{687aa}
\end{eqnarray}

Notice that the infrared regulator $\mu$, completely cancels in this correlation function.
Conversely, for the correlator
$\langle \sigma\mu\mu^\dagger\sigma\rangle$, for instance,
the $b^2$ term would have the sign of the first logarithm reversed, thus
producing an overall $\ln \mu$-factor that would force it to vanish.
The infrared regulator, therefore provides an efficient mechanism of enforcing the relevant selection rules.
The ultraviolet regulator $\epsilon$, appearing in the unphysical self-interaction terms, conversely,  may be removed by renormalizing, respectively, the 
operators $\sigma$ and $\mu$ in the correlation functions.

The natural choice for the bosonization of the Weyl fermions, therefore, is
\begin{eqnarray}
\Psi_L =\left ( \begin{array}{l}           
                                    \sigma \mu \\
                                    \sigma \mu^\dagger
                                        \end{array} 
                              \right )_L   \ \ \
\Psi_R =\left ( \begin{array}{l}           
                                    \sigma^\dagger \mu \\
                                    \sigma^\dagger \mu^\dagger
                                        \end{array} 
                              \right )_R   \ \ \
\label{690a}
\end{eqnarray}
where the $\sigma, \mu$ operators are expressed, respectively, in terms of the chiral tensor fields $B_{\mu\nu}^{L,R}$.
With the choice $a^2 + b^2 =6\pi^3$
we reproduce the correlation functions (\ref{684a}) completely within the framework of the bosonic
field theory.

The correlator (\ref{687aa}), in particular, would correspond, through bosonization, to $\langle \Psi_L(x)\Psi^\dagger_L(y) \rangle_{11}$.
One can easily verify that the remanining correlation functions in (\ref{684a}) are correctly reproduced by
the bosonization formulas above. Such formulas correspond in D=4 to the Mandelstam bosonization formula of D=2 \cite{sm}.

Using these bosonization formulas, we can obtain, for instance, a bosonized version of the Dirac particles mass term
(\ref{675abc}) that would befit the electron. This would be a generalization of the sine-Gordon theory.

Let us now take the dual operators at a constant time, namely,
\begin{eqnarray}
\sigma(S(C_{\textbf{y}}),t) &=& \exp \left \{ - i\frac{b}{2\pi \rho} \int_{S(C_{\textbf{y}})} d^2 \xi^{ij}B_{ij}(\mathbf{\xi},t)\right \}.
\nonumber  \\ 
\mu(\textbf{x},t) &=& \exp \left \{  ia \int_{-\infty}^\textbf{x} d \xi^i \Box^{-1/2}\epsilon^{ijk} \Pi^{jk}(\mathbf{\xi},t)\right \}.
\nonumber \\
\label{629}
\end{eqnarray}
and let us determine their commutation relations.
Using canonical equal-time commutation rules for the Kalb-Ramond field and its conjugate momentum $\Pi^{jk}$,
we obtain, for $\Psi_{L,1}(\textbf{x};C_{\textbf{x}}, t)=\sigma(C_{\textbf{x}}, t)\mu(\textbf{x}, t)$, for instance
\begin{eqnarray}
\Psi_{L,1}(\textbf{x};C_{\textbf{x}}, t)\Psi_{L,1}(\textbf{y};C_{\textbf{y}}, t) =\ \ \ \ \ \ \ \ \ \ \ \ \ \ \ \ \ \ \ \ \ \ \ \ \ \ \ \ 
\nonumber \\
\exp \left \{i ab \epsilon(2\pi - \Omega(\textbf{x};C_{\textbf{y}})) \right \}
\Psi_{L,1}(\textbf{y};C_{\textbf{y}}, t)\Psi_{L,1}(\textbf{x};C_{\textbf{x}}, t)
\nonumber \\
\label{634}
\end{eqnarray}
where $\Omega(\textbf{x};C_{\textbf{y}})$ is the solid angle comprised by the curve $C_{\textbf{y}}$ with respect to the point
$\textbf{x}$. Since, we want to describe local fields through the bosonization process, as mentioned before, we are
going to take the limit where the curve shrinks to a point and the solid angle, consequently reduces to zero. In this case,
the exponential factor above becomes a constant: $e^{iab}$. The only choice consistent  with multiple
commutations is, then, $ab=\pi$.This
reflects the fact that only fermion or boson local fields are allowed in D=4. For the string objects created by $\sigma(C)$,
however, the solid angle would not be zero and an arbitrary spin $s=\frac{ab}{2\pi}$ would be allowed \cite{em3}.

It is very instructive to investigate how the bosonized field behaves under a gauge transformation of the bosonic gauge
field, namely $B_{\mu\nu} \rightarrow B_{\mu\nu} +\partial_\mu \Lambda_\nu - \partial_\nu \Lambda_\mu$. We
immediately see that the operator $\mu$ is gauge invariant, whereas 
\begin{eqnarray}
\sigma(S(C_{\textbf{y}}),t) &\rightarrow& 
\sigma(S(C_{\textbf{y}}),t)\exp \left \{ - i\frac{b}{2\pi \rho} \oint_{C_{\textbf{y}}} d\xi^{i}\Lambda_{i}(\mathbf{\xi},t)\right \}
\nonumber \\
&\equiv& \sigma(S(C_{\textbf{y}}),t) e^{-i \varphi(y)}
\nonumber \\
\varphi(y)&=&\frac{b}{2\pi \rho} \oint_{C_{\textbf{y}}} d\xi^{i}\Lambda_{i}(\mathbf{\xi},t)
\label{629b}
\end{eqnarray}
We see that a gauge transformation of the bosonic tensor field emerges as an U(1) gauge transformation of the
Weyl fermions $\Psi_L$ and $\Psi_R$. The U(1) gauge transformation of a Dirac field, such as the electron field,
for instance, according to (\ref{690a}) and (\ref{677a}), would be obtained by simultaneous gauge transformations of the
chiral bosonic tensor fields $B^L_{\mu\nu}$ and $B^L_{\mu\nu}$ with opposite gauge parameters $\Lambda^R_\mu
= - \Lambda^L_\mu \equiv \Lambda_\mu$.
Then, the U(1) transformation of a Dirac field is such that
\begin{eqnarray}
\Psi_D &\rightarrow& e^{i \varphi(x)} \Psi_D
\nonumber \\
\varphi(x)&=&\lim_{\rho\rightarrow 0}\frac{b}{2\pi \rho} \oint_{C_{\textbf{x}}} d\xi^{i}\Lambda_{i}(\mathbf{\xi},t)
\label{629c}
\end{eqnarray}

{\textbf{VII.\,Magneto-Conductance}}

From (\ref{652ab})-(\ref{652ad}) we can infer that, in the presence of an applied external EM field, 
the chiral conductivity tensor will exhibit a term proportional to
\begin{eqnarray}
\delta^{ij}\left[ |\textbf{E}|^2|\textbf{A}|^2-(\textbf{E}\cdot\textbf{A})^2\right ]\propto\ \ \ \ \ \ \ \ \ \ \ \ \ \ \
\nonumber \\
\delta^{ij}\left[ |\textbf{E}|^2|\textbf{B}|^2+(\textbf{E}\cdot\textbf{B})^2
-   |\textbf{E}|^2(\hat{\textbf{r}}\cdot\textbf{B})^2-   |\textbf{B}|^2(\hat{\textbf{r}}\cdot\textbf{E})^2\right ]
\label{62c}
\end{eqnarray}
This will account for the magneto-conductance observed in Weyl semimetals \cite{nn,ss,weyl}.

{\textbf{VIII.\,Concluding Remarks}}

The bosonization of Dirac and Weyl fields in four-dimensional spacetime, obtained in the present work, opens several
new possibilities.
Under this new perspective, one may start to inquire about what ultimately are the so-called elementary particles such as
electrons, quarks and so on. What are their fundamental attributes, such as charge, spin, color, etc.
The bosonization of electrons and Weyl fermions, reported here, indicates that these elementary fermions are
topological excitations in the framework of a tensor gauge field theory. Electric charge, for instance becomes the
topological charge of the bosonic tensor theory. 

A remarkable connection emerges from our method. As we know, the electromagnetic interaction
results from imposing the invariance under U(1) gauge transformations. As we have seen, however,
these became the resulting effect
of an underlying gauge invariance of the tensor gauge field associated to the particles that carry the source of
the electromagnetic field.

Our results seem to indicate that the bosonic tensor gauge field is the underlying matrix upon which the fermionic matter fields are
created. The properties of these fermionic matter particles, such as the way they acquire mass, for instance, must be 
deeply influenced by the subjacent boson field. This may bring some light to the physics 
underlying the Higgs particle.

This work was partially supported by CNPq and FAPERJ.


\begin{thebibliography}{99}

\bibitem{weyl} S. -Y. Xu, $et$ $al.$, Science 347, 294 (2015); $ibid$ Science 349, 613 (2015);
                      B. Q. Lv $et$ $al.$ Phys. Rev. X5, 031013 (2015)

\bibitem{bos1d}  E. Lieb and D. C. Mattis, J. Math. Phys. 6, 304 (1965)

\bibitem{bos1d1} S. R. Coleman, Phys. D11, 2088 (1975)

\bibitem{sm} S. Mandelstam, Phys. D11, 3026 (1975)

\bibitem{tl} S. Tomonaga, Prog. Th. Phys. 5, 544 (1950);
                    J. M. Luttinger, J. Math. Phys. 4, 1154 (1963); 
                   F. D. M. Haldane, J. Phys. C 14, 2585 (1981)

\bibitem{ls} J. H. Lowenstein and J. A. Swieca, Ann. of Phys. 68, 172 (1971).

\bibitem{3} A. Luther, Phys. Rep. 49, 262 (1979)

\bibitem{4} F. Wilczek and A. Zee, Phys. Rev. Lett. 51, 2250 (1983);
                  J. Fr\" ohlich and P. A. Marchetti, Lett. Math. Phys. 16, 347 (1988);
                  A. M. Polyakov, Mod. Phys. Lett. A3, 325 (1988);
                  I. Dzyaloshinskii, A. M. Polyakov and P. Wiegmann, Phys. Lett. A127, 112 (1988);
                  S. N. Deser and A. N. Redlich, Phys. Rev. Lett. 61, 1541 (1988);
                  P. K. Panigrahi, S. Roy and W. Scherer,  Phys. Rev. Lett. 61, 2827 (1988);
                  L. Huerta and M. Ruiz-Altaba, Phys. Lett. B216, 371 (1989)

\bibitem{5} M. L\" uscher, Nucl. Phys. B326, 557 (1989)


\bibitem{bos4d1} E. Fradkin and F. A. Schaposnik, Phys. Lett. B338, 253 (1994) ;  
                           F. A. Schaposnik, Phys. Lett. B356, 253 (1995);
                           J. C. Le Guillou, C. N\' u\~ nez and F. A. Schaposnik, Ann. of Phys. 251, 426 (1996)
                           C. D. Fosco and F. A. Schaposnik, Phys. Lett. B391, 136 (1997)

\bibitem{bos4d}   C. P. Burgess and F. Quevedo, Nucl. Phys. B421, 373 (1994).
                              

\bibitem{bos4d2}  D. G. Barci, C. D. Fosco and L. D. Oxman, Phys. Lett. B375 , 267 (1996);
                          


\bibitem{em1} E. C. Marino, Phys. Lett. B263, 63 (1991).

\bibitem{odd} L. P. Kadanoff and H. Ceva, Phys. Rev. B3, 3918 (1971);
                      F. Wegner, J. Math. Phys, 12, 2259 (1971);
                      
\bibitem{odd1} G. 't Hooft, Nucl. Phys. B138, 1 (1978)

\bibitem{fm} J. Fr\" ohlich and P. A. Marchetti, Commun. Math. Phys. 116, 127 (1988);

\bibitem{em2} E. C. Marino and J. A. Swieca, Nucl. Phys. B170 [FS1], 175 (1980);
                      E. C. Marino, B. Schroer and J. A. Swieca, Nucl. Phys. B200 [FS4], 473 (1982)

\bibitem{bm} R. Banerjee and E. C. Marino, Nucl. Phys. B507, 501 (197);

\bibitem{gth}  G. 't Hooft, Phys. Rep. 142, 357 (1986)

\bibitem{chi} S. Adler, Phys. Rev. 177, 2496 (1969);  J. S. Bell and R. Jackiw, N. Cimento A60, 47 (1969)

\bibitem{nn} H. B. Nielsen and M. Ninomiya, Phys. Lett. 130, 389 (1983)

\bibitem{ss} D. T. Son and B. Z. Spivak, Phys. Rev. B88, 104412 (2013)

\bibitem{em3} E. C. Marino, Phys. Lett. B393, 383 (1997)

\end{thebibliography}
\end{document}